\documentclass[12pt]{article}
\usepackage{amsmath, amsthm, amssymb, amsfonts, enumerate,bm,url}
\usepackage[figuresright]{rotating}
\usepackage{lscape}
\usepackage{multirow}
\usepackage{natbib}

\begin{document}
\newcommand{\abs}[1]{\left\vert#1\right\vert}
\newcommand{\set}[1]{\left\{#1\right\}}
\newcommand{\eps}{\varepsilon}
\newcommand{\To}{\rightarrow}
\newcommand{\inv}{^{-1}}
\newcommand{\ihat}{\hat{\imath}}
\newcommand{\var}{\mbox{Var}}
\newcommand{\sd}{\mbox{SD}}
\newcommand{\cov}{\mbox{Cov}}
\newcommand{\f}{\frac}
\newcommand{\fI}[1]{\frac{1}{#1}}
\newcommand{\what}[1]{\widehat{#1}}
\newcommand{\hhat}[1]{\what{\what{#1}}}
\newcommand{\wtilde}[1]{\widetilde{#1}}
\newcommand{\bdot}{\bm{\cdot}}
\newcommand{\Th}{\theta}
\newcommand{\qmq}[1]{\quad\mbox{#1}\quad}
\newcommand{\qm}[1]{\quad\mbox{#1}}
\newcommand{\mq}[1]{\mbox{#1}\quad}
\newcommand{\tr}{\mbox{tr}}
\newcommand{\logit}{\mbox{logit}}
\newcommand{\noi}{\noindent}
\newcommand{\bni}{\bigskip\noindent}
\newcommand{\bul}{$\bullet$ }
\newcommand{\bias}{\mbox{bias}}
\newcommand{\conv}{\mbox{conv}}
\newcommand{\spn}{\mbox{span}}
\newcommand{\colspace}{\mbox{colspace}}
\newcommand{\mF}{\mathcal{F}}
\newcommand{\mI}{\mathcal{I}}
\newcommand{\mL}{\mathcal{L}}
\newcommand{\mP}{\mathcal{P}}
\newcommand{\mS}{\mathcal{S}}
\newcommand{\mT}{\mathcal{T}}
\newcommand{\mX}{\mathcal{X}}
\newcommand{\bbR}{\mathbb{R}}
\newcommand{\fwer}{\mbox{FWE}}
\newcommand{\fweI}{FWE$_I$}
\newcommand{\fweII}{FWE$_{II}$}
\newcommand{\fdr}{\mbox{FDR}}
\newcommand{\fnr}{\mbox{FNR}}
\newcommand{\pfdr}{\mbox{pFDR}}
\newcommand{\pfnr}{\mbox{pFNR}}
\newcommand{\mte}{\mbox{MTE}}
\newcommand{\vphi}{\varphi}
\newcommand{\Bern}{\mbox{Bern}}

\newtheorem{theorem}{Theorem}[section]
\newtheorem{corollary}{Corollary}[section]
\newtheorem{conjecture}{Conjecture}[section]
\newtheorem{proposition}{Proposition}[section]
\newtheorem{lemma}{Lemma}[section]
\newtheorem{definition}{Definition}[section]
\newtheorem{example}{Example}[section]
\newtheorem{remark}{Remark}[section]

\title{{\bf\Large Asymptotically optimal sequential FDR and pFDR control with (or without) prior information on the number of signals}}

\author{\textsc{Xinrui He}$^\dag$ and \textsc{Jay Bartroff}$^\ddag$\footnote{Corresponding author. Email: \textsf{bartroff@usc.edu}}\\
\small{$^\dag$Acumen LLC and the SPHERE Institute, Burlingame, California, USA}\\ 
\small{$^\ddag$Department of Mathematics, University of Southern California, Los Angeles, California, USA}\\ 
}  

\date{}
\maketitle

\abstract{We investigate asymptotically optimal multiple testing procedures for streams of sequential data in the context of prior information on the number of false null hypotheses (``signals''). We show that the ``gap'' and ``gap-intersection'' procedures, recently proposed and shown by \citet[][\textit{Electron.\ J.\ Statist.}]{Song17} to be asymptotically optimal for controlling type~1 and 2 familywise error rates (FWEs), are also asymptotically optimal for controlling FDR/FNR when their critical values are appropriately adjusted. Generalizing this result, we show that these procedures, again with appropriately adjusted critical values, are asymptotically optimal for controlling \emph{any} multiple testing error metric that is bounded between multiples of FWE in a certain sense.  This class of metrics includes FDR/FNR but also pFDR/pFNR, the per-comparison and per-family error rates, and the false positive rate. Our analysis includes asymptotic regimes in which the number of null hypotheses approaches $\infty$ as the type~1 and 2 error metrics approach $0$.} 

\section{Introduction}
For decades, the problem of how to efficiently and powerfully test  multiple statistical hypotheses while controlling some notion of type~1 error frequency has been fundamental and active in the statistics methodology literature. The majority of this research has concerned testing procedures which operate on fixed-sample data, typically in the form of a collection of $p$-values, one for each null hypothesis, which are combined in some way to reach reject/accept decisions for each null. Recently, driven by applications where data is streaming or arrives sequentially, multiple testing procedures that can handle sequential data have been proposed and studied.  Applications with data of this type include the analysis of streaming internet data \citep{Wegman03}, multiple channel signal detection in sensor networks \citep{Dragalin99,Mei08}, high throughput sequencing technology \citep{Jiang12}, and multi-arm and multiple endpoint clinical trials \citep{Bartroff10e}.

Existing multiple testing procedures for sequential data occur in essentially two forms.  In one, the individual data streams can be terminated at different times \citep[e.g.,][]{Bartroff14b,Bartroff15c,Bartroff18,Malloy14}. In the other form, termination of sampling must occur at the same time for all streams \citep[e.g.,][]{De12,De12b,Song17,Song19}. The applications mentioned in the previous paragraph span both of these forms. Recently, a third form of ``sequential'' multiple testing procedure has been studied in which not the data, but rather the hypotheses themselves arrive sequentially in time, each with its own fixed-sample $p$-value. \citet{Javanmard18} proposed an FDR-controlling procedure in this setup, which \citet{Chen17} showed to be optimal under certain distributional assumptions. A recent manuscript by \citet{Zrnic18} studies optimality more generally in this setup.

The current paper adopts the second form described above in which all the data streams are terminated at the same time, and we investigate the optimal choice of that stopping rule subject to desired bounds on the type~1 and 2 error metrics, asymptotically as these bounds approach zero at arbitrary rates. This is done under the condition of prior information on the number of false null hypotheses (``signals'') in the form of a known number of signals, or known bounds on this number. The latter case of known bounds on the number of signals includes the non-informative setting with lower bound $0$ and upper bound equal to the total number of null hypotheses.\footnote{This case is the reason for the word ``without'' in the paper's title.}  Our work springs from and generalizes the results of \citet{Song17} who found asymptotically optimal procedures in these settings when the error metrics are type~1 and 2 familywise error rates (FWEs). By modifying the procedures of Song and Fellouris, we find the corresponding asymptotically optimal procedures for controlling the false discovery rate (FDR) and its type~2 analog, the false non-discovery rate (FNR). Further, we are able to find the asymptotically optimal procedures for controlling \emph{any} multiple testing error metric that is bounded between multiples of FWE in a certain sense, which includes FDR/FNR, the positive false discovery and non-discovery rates (pFDR and pFNR), the per comparison error rate, and other metrics.  Further, we are able to consider asymptotic regimes in which the number~$J$ of null hypotheses approaches $\infty$.  

After introducing notation and describing our approach in Section~\ref{sec:setup}, the case of the number of signals known exactly is addressed in Section~\ref{sec:m} where first the general result for arbitrary error metrics is stated, followed by its application to FDR/FNR and pFDR/pFNR. The case of bounds on the number of signals is addressed in Section~\ref{sec:bds} where, again, the general result for arbitrary error metrics followed by its application to FDR/FNR and pFDR/pFNR.  Simulation studies of procedures for FDR/FNR control in  finite-sample settings are presented in Section~\ref{sec:sim}, and we conclude with a discussion of related issues in Section~\ref{sec:disc}. 

\section{Set up and summary of our approach}\label{sec:setup}

\subsection{Notation and assumptions}\label{sec:not}
We consider $J\ge 2$ independent data streams
\begin{equation*}
X^j=\{X_1^j, X_2^j,\ldots\},\quad j\in[J],
\end{equation*} where $[J]$ denotes $\{1,2,\ldots, J\}$ throughout.  We assume throughout the paper that the streams are independent of each other, but not always that the elements of a given stream $X_1^j, X_2^j,\ldots$ are independent. In particular, we will prove error control of the proposed procedures only under independence of the streams, but make the additional assumption that each stream is made up of i.i.d.\ observations in order to prove asymptotic optimality.

Letting $P^j$ denote the probability distribution of $X^j$, we consider simultaneously testing $J$ null-alternative simple hypothesis pairs
\begin{equation}\label{hyps}
H_0^j: P^j=P_0^j\qmq{vs.}H_1^j: P^j=P_1^j,\quad j\in[J],
\end{equation} where, for each $j$, the  $P_0^j$ and $P_1^j$ are known and distinct distributions, assumed to be mutually absolutely continuous when restricted to $\sigma_n=\sigma(\{X_i^j:\; i\in[n],\; j\in[J]\})$,  the $\sigma$-field generated by the observations 
\begin{equation}\label{obs.time.n}
X_1^j,X_2^j,\ldots, X_n^j,\quad j\in[J].
\end{equation}
We utilize the informal but illustrative terminology that refers to the $j$th stream as \textit{noise} (resp.\ \textit{signal}) if $H_0^j$ (resp.\ $H_1^j$) is true. We shall refer to the subscript~$n$ in $X_n^j$ as \textit{time} and assume that at time~$n$, the observations~\eqref{obs.time.n} (and \emph{only} these observations) have been observed.  We define a \textit{sequential test} (or \textit{procedure}) of \eqref{hyps} as a pair of random variables $(T,D)$ such that $T$ is a $\{\sigma_n\}$-stopping time (i.e., the event $\{T=n\}\in\sigma_n$) and $D=(D^1,\ldots, D^J)$ is a $\sigma_T$-measurable decision rule taking values in $\{0,1\}^J$. The interpretation of $(T,D)$ is that sampling of all streams is terminated at time~$T$ and  the procedure classifies the $j$th stream as noise (resp.\ signal) if $D^j=0$ (resp.\ $D^j=1$), for each $j\in[J]$.

For each pair of distributions $P_0^j$ and $P_1^j$ in \eqref{hyps}, let $f_0^j$ and $f_1^j$ denote the respective density functions with respect to some common measure~$\mu^j$.  Denote the log-likelihood ratio for the $j$th stream at time $n$ by
\begin{equation*}
\lambda^j(n):=\log \frac{f_1^j(X_1^j,\ldots, X_n^j)}{f_0^j(X_1^j,\ldots, X_n^j)}
\end{equation*}
and the Kullback-Leibler information numbers by
 \begin{equation}\label{KL.inf.def}
I_0^j:=\int \log\lambda^j(1)^{-1} f_0^j(X_1^j)d\mu^j\qmq{and}I_1^j:=\int \log\lambda^j(1) f_1^j(X_1^j)d\mu^j.
\end{equation} Define
 \begin{equation}\label{KL.var}
v_0^j:=\int \left(\log\lambda^j(1)^{-1}-I_0^j\right)^2 f_0^j(X_1^j)d\mu^j\qmq{and}v_1^j:=\int \left(\log\lambda^j(1) -I_1^j\right)^2f_1^j(X_1^j)d\mu^j.
\end{equation}

We assume the likelihood ratio statistics satisfy 
\begin{equation}\label{LLN}
P_0^j\left(\lim_{n\To\infty}\lambda^j(n)=-\infty\right) = P_1^j\left(\lim_{n\To\infty}\lambda^j(n)=\infty\right)=1
\end{equation}
 for all $j\in[J]$. For example, if the elements of the stream $X_1^j,X_2^j,\ldots$ are i.i.d., then \eqref{LLN} follows from the strong law of large numbers. Since, as mentioned in the first paragraph of this section, we assume the streams are independent throughout but only make the i.i.d.\ assumption for some of our results, the assumption \eqref{LLN} guarantees that the proposed sequential tests  will terminate a.s.\ even when the i.i.d.\ assumption is not made. It is under the i.i.d.\ assumption that the Kullback-Leibler information numbers \eqref{KL.inf.def} are utilized.

 In what follows we will make frequent use of the log-likelihood ratios' order statistics which we denote by
  \begin{equation*}
\lambda^{(1)}(n)\ge \lambda^{(2)}(n)\ge\ldots\ge \lambda^{(J)}(n),
\end{equation*}
with ties broken arbitrarily.  To refer to a particular order statistic let $i_j(n)$, $j\in[J]$, be such that $\lambda^{i_j(n)}(n)=\lambda^{(j)}(n)$. Also, letting $|\cdot|$ denote set cardinality throughout, define $p(n):=\left|\{j\in[J]:\; \lambda^j(n)>0\}\right|$ to be the number of positive $\lambda^j$ at time $n$.

We will describe states of nature by subsets $A\subseteq [J]$ which we call \textit{signal sets}, describing the situation where $H_0^j$ is false (signal) for all $j\in A$, and $H_0^j$ is true (noise) for all $j\in A^c:=[J]\setminus A$. For a signal set~$A$, let $P_A$ denote the probability measure determined by $A$, i.e.,
\begin{equation}\label{P_A}
P_A := \bigotimes_{j = 1}^J P_{\bm{1}\{j\in A\}}^j,
\end{equation}
where $\bm{1}\{\cdot\}$ denotes an indicator function throughout. Let $E_A$ denote expectation with respect to $P_A$.  For a given signal set~$A$, let
\begin{equation*}
\eta_0^A := \min_{j\in A^c} I_0^j\qmq{and} \eta_1^A := \min_{j\in A} I_1^j,
\end{equation*}
which can be thought of as the ``worst case'' information numbers for signal set~$A$, and appear in the expected sample sizes of optimal procedures.

For any multiple testing procedure under consideration, let $V$ denote the number of true null hypotheses rejected (i.e., the number of false positives), $W$ the number of false null hypotheses accepted (i.e., the number of false negatives), and $R$ the number of null hypotheses rejected. The number of null hypotheses accepted is therefore $J-R$. Under signal set $A$  the type~1 and 2 familywise error rates (FWEs)  are
\begin{equation*}
\fwer_{1,A}=P_A(V\ge 1),\quad \fwer_{2,A}=P_A(W\ge 1)
\end{equation*}
and the false discovery and non-discovery rates (FDR, FNR) are
\begin{equation*}
\fdr_A=E_A\left(\frac{V}{R\vee 1}\right),\quad \fnr_A=E_A\left(\frac{W}{(J-R)\vee 1}\right),
\end{equation*}
where $x\vee y=\max\{x,y\}$. In these and other multiple testing metrics, we will include the procedure being evaluated as an argument (e.g., $\fwer_{1,A}(T,D)$) when needed but omit it when it is clear from the context or when a statement holds for any procedure. Likewise, we will omit the subscript $A$ in expressions that hold for arbitrary signal sets, such as in the next paragraph.

\subsection{Summary of our approach}
For the asymptotic theory governing optimal FDR and FNR control it will suffice to consider the following elementary bounds between these metrics and the type~1 and 2 FWEs. On one hand,
\begin{equation}\label{fdr<fwe}
\fdr=E\left(\frac{V}{R\vee 1}\right)= E\left(\frac{V}{R\vee 1}\cdot \bm{1}\{V\ge 1\}\right) \le E\left(1\cdot\bm{1}\{V\ge 1\}\right)=P(V\ge 1) = \fwer_1.
\end{equation}
On the other hand,
\begin{equation}\label{fdr>fwe}
\fdr=E\left(\frac{V}{R\vee 1}\cdot\bm{1}\{V\ge 1\}\right)\ge E\left(\frac{1}{J}\cdot\bm{1}\{V\ge 1\}\right)= \frac{1}{J}\cdot P(V\ge 1)= \frac{1}{J}\cdot\fwer_1.
\end{equation}
Similar arguments show that
\begin{equation}\label{fnr<fwe}
\frac{1}{J}\cdot\fwer_{2}\le \fnr \le \fwer_{2}.
\end{equation}
Although crude, it will turn out that bounds like \eqref{fdr<fwe}-\eqref{fnr<fwe} suffice to show that  sequential procedures that are asymptotically optimal for FWE control are also asymptotically optimal for FDR/FNR control. More generally, we will show that the sequential procedures that are asymptotically optimal for FWE control are also asymptotically optimal (with slightly modified critical values) for control of \emph{any}  multiple testing error metric whose  type~1 and 2 versions are bounded above by some constant multiple of the corresponding $\fwer_i$ when evaluated on the optimal procedure, and are bounded below by some constant multiple of the corresponding $\fwer_i$ when evaluated on any procedure; these statements are made precise in conditions (\ref{m.mte<})-(\ref{m.mte>}) of Theorems~\ref{thm:gap} and \ref{thm:GI}.  Furthermore, it will turn out that these ``bounds'' between the error metrics need not be constant and the lower bounds can be allowed to approach $0$ and the upper bounds $\infty$, allowing asymptotic regimes in which the number~$J$ of null hypotheses approaches $\infty$.  See part~(\ref{m.asymp}) of Theorems~\ref{thm:gap} and \ref{thm:GI}.

To state these more general results, we denote the type 1 and 2 versions of a generic multiple testing error metric by $\mte=(\mte_1,\mte_2)$, which is any pair of functions mapping multiple testing procedures into $[0,1]$. As above, we will add a signal set~$A$ as a subscript and a procedure $(T,D)$ as an argument when needed. Our more general results in Theorems \ref{thm:gap} and \ref{thm:GI} will produce asymptotic optimality results for FDR/FNR in Corollaries~\ref{cor:gap.fdr} and \ref{cor:GI.fdr} and for pFDR/pFNR in Corollaries~\ref{cor:gap.pfdr} and \ref{cor:GI.pfdr}, after verifying that pFDR/pFNR satisfy similar bounds.

We will use $\mP$ to denote a \textit{signal class} which is a collection of signal sets; more precisely, $\mP$ is a subset of the power set of $[J]$.  We will consider the classes of sequential tests~$(T,D)$ controlling the type~1 and 2 versions of various multiple testing error metrics at specific levels\footnote{Here we do not explicitly require that $\alpha,\beta<1$ in the definitions of the classes that follow. This is because below we will multiply and divide inputs $\alpha,\beta$ to the class \eqref{Del.MTE} by various positive constants. This does not present a problem because, for example, if $\alpha\ge 1$ then the class requirement in $\Delta_\mP^{\tiny{\mte}}(\alpha,\beta)$ that $\mte_{1,A}\le\alpha$ will  automatically be satisfied since $\mte_i\in[0,1]$ by assumption. Similar statements apply if $\beta\ge 1$.}~$\alpha,\beta>0$. For a generic metric MTE, let
\begin{equation}\label{Del.MTE}
\Delta_\mP^{\tiny{\mte}}(\alpha,\beta)=\{(T,D): \mbox{$\mte_{1,A}\le\alpha$ and $\mte_{2,A}\le\beta$ for all $A\in\mP$}\}.
\end{equation}
When the MTE is FWE we have
\begin{equation*}
\Delta_\mP^{\tiny{\fwer}}(\alpha,\beta)=\{(T,D): \mbox{$\fwer_{1,A}\le\alpha$ and $\fwer_{2,A}\le\beta$ for all $A\in\mP$}\}
\end{equation*}
and, by a slight but obvious abuse of notation, for FDR/FNR control we have
\begin{equation*}
\Delta_\mP^{\tiny{\fdr}}(\alpha,\beta)=\{(T,D): \mbox{$\fdr_{A}\le\alpha$ and $\fnr_{A}\le\beta$ for all $A\in\mP$}\}.
\end{equation*}

\section{Number of signals known exactly}\label{sec:m}
For $m\in[J-1]$ let $\mP_m=\{A\subseteq [J]: |A|=m\}$. Thus $\mP_m$ is the class of signal sets with exactly $m$ signals. For a given such $m$, \citet{Song17} defined the following \textit{gap rule} $(T_G(c),D_G(c))$ with threshold~$c>0$, which is the sequential procedure that stops sampling as soon as  the ``gap'' between the $m$th and $(m + 1)$st ordered log-likelihood ratio statistics is at least $c$. The decision taken at that time is that the null hypotheses with the largest $m$ statistics contain signal. That is,
\begin{align}
T_G(c) &:= \inf\{n \geq 1\ :\ \lambda^{(m)}(n) - \lambda^{(m + 1)}(n)\ \geq\ c\},\label{gapT}\\
D_G(c) &:= \{i_1(T_G(c)),...,i_m(T_G(c))\}.\label{gapD}
\end{align}

Theorem~\ref{thm:gap} establishes admissability and asymptotic optimality of the gap rule for control of any metric whose type~1 version is bounded above by some constant multiple of $\fwer_1$ when evaluated on the gap rule in the sense of \eqref{m.mte<disp}, and whose type~2 version is is bounded above by some constant multiple of $\fwer_2$ when evaluated on any procedure in the sense of \eqref{m.mte>ineq}.

\subsection{Main result and its application to FDR/FNR and pFDR/pFNR control}\label{sec:gen.m}
\begin{theorem}\label{thm:gap} 
Fix $m\in[J-1]$ and let $(T_G(c),D_G(c))$ denote the gap rule with number of signals~$m$ and threshold~$c>0$. Let MTE be a multiple testing error metric such that:
\begin{enumerate}[(i)]
\item\label{m.mte<}  there is a constant~$C_1$ such that
\begin{equation}\label{m.mte<disp}
\mte_{i,A}(T_G(c),D_G(c)) \le C_1\cdot \fwer_{i,A}(T_G(c),D_G(c))
\end{equation}
for $i=1$ and $2$, for all $A\in\mP_m$, and for all $c>0$, and
\item\label{m.mte>} there is a constant~$C_2$ such that
\begin{equation}\label{m.mte>ineq}
\mte_{i,A}(T,D) \ge C_2\cdot \fwer_{i,A}(T,D)
\end{equation}
for $i=1$ and $2$, for all $A\in\mP_m$, and for all procedures $(T,D)$.
\end{enumerate}
Given $\alpha,\beta\in(0,1)$ let $(T_G^*,D_G^*)$ denote the gap rule with number of signals~$m$ and threshold
\begin{equation*}
c =|\log((\alpha/C_1) \wedge (\beta/C_1))|  + \log(m(J - m)).
\end{equation*}
 Then the following hold.
\begin{enumerate}[(1)]
\item\label{m.adm} Under the assumption that the streams $X^1,\ldots, X^J$ are independent, $(T_G^*,D_G^*)$ is admissible for MTE control. That is,  
\begin{equation}\label{T*inD}
(T_G^*,D_G^*)\in \Delta_{\mP_m}^{\tiny{\mte}}(\alpha,\beta).
\end{equation}
\item\label{m.opt} Under the additional assumption that each stream $X^j=(X_1^j, X_2^j,\ldots)$ is made up of i.i.d.\ random variables~$X_1^j, X_2^j,\ldots$, the procedure $(T_G^*,D_G^*)$ is asymptotically optimal for MTE control with respect to class~$\mP_m$. That is, for all $A \in \mP_m$,
\begin{equation}\label{m.ET*.opt}
E_A(T_G^*) \sim \frac{|\log(\alpha \wedge \beta)|}{\eta_1^A + \eta_0^A} \sim \inf_{(T, D)\in \Delta_{\mP_m}^{\tiny{\mte}}(\alpha,\beta)}E_A(T)
\end{equation}
as $\alpha,\beta\To 0$.
\item\label{m.asymp} Parts~(\ref{m.adm}) and (\ref{m.opt}) still hold if $J, m, C_1$, and $C_2$ are allowed to vary with $\alpha, \beta$ as long as 
\begin{equation}\label{m.J.inf}
J=o(|\log(\alpha\wedge \beta)|^{1/4}),\quad \log C_1=o(|\log(\alpha\wedge \beta)|),\qmq{and} \log C_2^{-1}=o(|\log(\alpha\wedge \beta)|),\end{equation} 
and 
\begin{gather}
\mbox{the information numbers $\{I_0^j, I_1^j: j\in[J]\}$ in \eqref{KL.inf.def} are bounded above $0$, and }\label{KL.inf.bound}\\
\mbox{the variances $\{v_0^j, v_1^j: j\in[J]\}$ in \eqref{KL.var} are bounded below $\infty$.}\label{KL.var.bound}
\end{gather}
 In particular, the asymptotic optimality in part~(\ref{m.opt}) still holds if $J, C_1\To\infty$ and $C_2\To 0$ as $\alpha,\beta\To 0$ as long as \eqref{m.J.inf}-\eqref{KL.var.bound} hold, with $m\in[J-1]$ allowed to vary arbitrarily.
\end{enumerate}
\end{theorem}

\bigskip

\noindent\textbf{Proof.} For part~(\ref{m.adm}), fix arbitrary $A\in\mP_m$. \citet[][Theorem~3.1]{Song17} establish that
\begin{equation*}
\fwer_{1,A}(T_G^*,D_G^*)\le \alpha/C_1\qmq{and}\fwer_{2,A}(T_G^*,D_G^*)\le \beta/C_1.
\end{equation*}
Applying \eqref{m.mte<disp} yields
$$\mte_{1,A}(T_G^*,D_G^*)\le C_1\cdot \fwer_{1,A}(T_G^*,D_G^*)\le C_1(\alpha/C_1)=\alpha.$$
A similar argument shows that $\mte_{2,A}(T_G^*,D_G^*)\le\beta$, establishing \eqref{T*inD}.

For part~(\ref{m.opt}), fix arbitrary $A\in\mP_m$ and consider $\alpha,\beta\To 0$. Let $\kappa$ denote $|\log(\alpha \wedge \beta)|\To\infty$. The first inequality in the following is established by \citet[][Lemma~5.2]{Song17}, and the rest are straightforward calculations: 
\begin{equation}\label{ETG.low}
E_A(T_G^*)\le \frac{c}{\eta_1^A + \eta_0^A}+O\left(m(J-m)\sqrt{c}\right) =\frac{\kappa}{\eta_1^A + \eta_0^A} + O(\log C_1)+O(J^2\sqrt{\kappa})\\
\sim \frac{\kappa}{\eta_1^A + \eta_0^A}.
\end{equation}
To establish that the last expression in \eqref{ETG.low} is also a lower bound for any procedure in $\Delta_{\mP_m}^{\tiny{\mte}}(\alpha,\beta)$, it follows from \eqref{m.mte>ineq} that $$\Delta_{\mP_m}^{\tiny{\mte}}(\alpha,\beta)\subseteq \Delta_{\mP_m}^{\tiny{\fwer}}(\alpha/C_2, \beta/C_2),$$ thus
$$\inf_{(T, D)\in \Delta_{\mP_m}^{\tiny{\mte}}(\alpha,\beta)}E_A(T) \ge \inf_{(T, D)\in \Delta_{\mP_m}^{\tiny{\fwer}}(\alpha/C_2, \beta/C_2)}E_A(T).$$
\citet[][Theorem~5.3]{Song17} establish that the latter is of order
\begin{equation}\label{m.TG.low}
\frac{|\log((\alpha/C_2)\wedge (\beta/C_2))|}{\eta_1^A + \eta_0^A} = \frac{\kappa}{\eta_1^A + \eta_0^A}-O(\log C_2^{-1}) \sim \frac{\kappa}{\eta_1^A + \eta_0^A}.
\end{equation}
Thus
\begin{equation}\label{m.inf.T.low}
\inf_{(T, D)\in \Delta_{\mP_m}^{\tiny{\mte}}(\alpha,\beta)}E_A(T) \ge \frac{\kappa}{\eta_1^A + \eta_0^A}(1+o(1))
\end{equation}
and combining this with \eqref{ETG.low} gives the desired result.

For part~(\ref{m.asymp}), letting $J, m, C_1$, and $C_2$ possibly vary with $\alpha,\beta$ such that \eqref{m.J.inf}-\eqref{KL.var.bound} hold does not affect the calculations for part~(\ref{m.adm}) since they are non-asymptotic.  The condition \eqref{KL.inf.bound} guarantees that the term $1/(\eta_1^A + \eta_0^A)$ in \eqref{m.ET*.opt} is bounded away from $\infty$, and it is bounded away from $0$ since the $\eta_i^A$ are minima of sets of finite numbers. A more explicit form of the $O(\cdot)$ term after the first inequality in \eqref{ETG.low} is
\begin{equation}\label{bigO.expl}
O\left(m(J-m)\sqrt{c\max_{i,j} v_i^j/(I_i^j)^3}\right),
\end{equation} which follows from results in renewal theory; see \citet[][Chapter~3.9]{Gut09} and \citet[][Theorem~2]{Mei08}. The conditions \eqref{KL.inf.bound}-\eqref{KL.var.bound} thus guarantee that this maximum is $O(1)$, thus \eqref{ETG.low} still holds when $J$ and $C_1$ satisfy \eqref{m.J.inf}, hence the same asymptotic upper bound on $E_A(T_G^*)$ holds. Similarly, \eqref{m.TG.low} still holds because $C_2$ satisfies \eqref{m.J.inf}, hence the same asymptotic lower bound~\eqref{m.inf.T.low} on $E_A(T)$ holds.\qed

\subsubsection{FDR/FNR control under known number of signals}\label{sec:gap.fdr}

To apply Theorem~\ref{thm:gap} to  FDR/FNR, we see that the upper bounds on  FDR/FNR in \eqref{fdr<fwe} and \eqref{fnr<fwe} suffice\footnote{In fact, \eqref{fdr<fwe} and \eqref{fnr<fwe} are stronger than what is needed since they hold for all procedures, whereas  condition~(\ref{m.mte<}) just requires this to hold for the gap rule.} for condition~(\ref{m.mte<}) of the theorem with $C_1=1$, and the lower bounds in \eqref{fdr>fwe} and \eqref{fnr<fwe} suffice for condition~(\ref{m.mte>}) with $C_2=1/J$, yielding the following corollary. And letting $J\To \infty$ such that $J=o(|\log(\alpha\wedge \beta)|^{1/4})$ satisfies the requirements in \eqref{m.J.inf} of both $J$ and $C_2=1/J$.

\begin{corollary}\label{cor:gap.fdr} For $m\in[J-1]$ and $\alpha,\beta\in(0,1)$, let $(T_G^*,D_G^*)$ denote the gap rule with number of signals~$m$ and threshold~$c =|\log(\alpha \wedge \beta)|  + \log(m(J - m))$. 
\begin{enumerate}[(1)]
\item\label{m.fdr.adm} Under the assumption that the streams $X^1,\ldots, X^J$ are independent, $(T_G^*,D_G^*)$ is admissible for FDR/FNR control. That is,  
$$(T_G^*,D_G^*)\in \Delta_{\mP_m}^{\tiny{\fdr}}(\alpha,\beta).$$
\item\label{m.fdr.opt} Under the additional assumption that each stream $X^j=(X_1^j, X_2^j,\ldots)$ is made up of i.i.d.\ random variables~$X_1^j, X_2^j,\ldots$, the procedure~$(T_G^*,D_G^*)$ is asymptotically optimal for FDR/FNR control with respect to class~$\mP_m$. That is, for all $A \in \mP_m$,
$$
E_A(T_G^*) \sim \frac{|\log(\alpha \wedge \beta)|}{\eta_1^A + \eta_0^A} \sim \inf_{(T, D)\in \Delta_{\mP_m}^{\tiny{\fdr}}(\alpha,\beta)}E_A(T)
$$
as $\alpha,\beta\To 0$.
\item Parts~(\ref{m.fdr.adm}) and (\ref{m.fdr.opt}) still hold if $J\To\infty$ as $\alpha, \beta\To 0$ such that $J=o(|\log(\alpha\wedge \beta)|^{1/4})$, \eqref{KL.inf.bound}-\eqref{KL.var.bound} hold, and $m\in[J-1]$ varies arbitrarily with $\alpha, \beta$.
\end{enumerate}
\end{corollary}

\subsubsection{pFDR/pFNR control under known number of signals}\label{sec:gap.pFDR}

Continuing to use  the notation of Section~\ref{sec:setup}, pFDR and its type~2 analog pFNR are defined as
\begin{equation}\label{pFDR.defs}
\pfdr=E\left(\left.\frac{V}{R}\right|R\ge 1\right)\qmq{and} \pfnr=E\left(\left.\frac{W}{J-R}\right|J-R\ge 1\right).
\end{equation} 
As above, we will add the signal set $A$ in the subscript and a procedure as an argument when needed. For $\alpha,\beta>0$ and signal class~$\mP$, define
$$\Delta_\mP^{\tiny{\pfdr}}(\alpha,\beta)=\{(T,D): \mbox{$\pfdr_{A}\le\alpha$ and $\pfnr_{A}\le\beta$ for all $A\in\mP$}\}.$$

To apply Theorem~\ref{thm:gap} to  pFDR/pFNR, unlike with FDR/FNR it is not possible to provide  universal upper bounds like \eqref{fdr<fwe} and \eqref{fnr<fwe}. Proceeding similarly, one obtains
\begin{multline}\label{pFDR<}
\pfdr=E\left(\left.\frac{V}{R}\right|R\ge 1\right) = E\left(\left.\frac{V}{R}\right|R\ge 1, V\ge 1\right)P(V\ge 1|R\ge 1)\le 1\cdot \frac{P(V\ge 1)}{P(R\ge 1)}\\
=\frac{\fwer_1}{P(R\ge 1)}.
\end{multline}
Similarly, $\pfnr\le P(W\ge 1)/P(J-R\ge 1)$, and because $P(R\ge 1)$ and $P(J-R\ge 1)$ cannot be bounded below in general, universal upper bounds like \eqref{fdr<fwe} and \eqref{fnr<fwe} do not hold for pFDR/pFNR. However, condition~(\ref{m.mte<}) in Theorem~\ref{thm:gap} merely requires such upper bounds to hold for the gap rule~$(T_G,D_G)$ with $m$ signals, which always rejects exactly $m$ nulls, and recalling that $1\le m\le J-1$ by assumption (i.e., $m\in[J-1]$), the gap rule satisfies $$P(R\ge 1)=P(J-R\ge 1)=1.$$ Thus, \eqref{pFDR<} yields $\pfdr(T_G,D_G)\le \fwer_1$ and $\pfnr(T_G,D_G)\le \fwer_2$, so condition~(\ref{m.mte<}) of the theorem holds with $C_1=1$. The lower bounds are more straightforward:
\begin{multline}\label{pFD/FW>}
\pfdr= E\left(\left.\frac{V}{R}\right|R\ge 1, V\ge 1\right)P(V\ge 1|R\ge 1)\ge \frac{1}{J}\cdot \frac{P(V\ge 1)}{P(R\ge 1)}\ge \frac{1}{J}\cdot \frac{\fwer_1}{1} \\
= \frac{1}{J}\cdot  \fwer_1,
\end{multline}
 and similarly 
 \begin{equation}\label{pFN/FW>}
\pfnr\ge (1/J)\cdot \fwer_2.
\end{equation}
Thus, condition~(\ref{m.mte>}) of the theorem is satisfied with $C_2=1/J$, yielding the following corollary.

\begin{corollary}\label{cor:gap.pfdr} For $m\in[J-1]$ and $\alpha,\beta\in(0,1)$, let $(T_G^*,D_G^*)$ denote the gap rule with number of signals~$m$ and threshold~$c =|\log(\alpha \wedge \beta)|  + \log(m(J - m))$. 
\begin{enumerate}[(1)]
\item\label{m.pdfr.adm} Under the assumption that the streams $X^1,\ldots, X^J$ are independent, $(T_G^*,D_G^*)$ is admissible for pFDR/pFNR control. That is,  
$$(T_G^*,D_G^*)\in \Delta_{\mP_m}^{\tiny{\pfdr}}(\alpha,\beta).$$
\item \label{m.pfdr.opt} Under the additional assumption that each stream $X^j=(X_1^j, X_2^j,\ldots)$ is made up of i.i.d.\ random variables~$X_1^j, X_2^j,\ldots$, the procedure~$(T_G^*,D_G^*)$ is asymptotically optimal for pFDR/pFNR control with respect to class~$\mP_m$. That is, for all $A \in \mP_m$,
$$
E_A(T_G^*) \sim \frac{|\log(\alpha \wedge \beta)|}{\eta_1^A + \eta_0^A} \sim \inf_{(T, D)\in \Delta_{\mP_m}^{\tiny{\pfdr}}(\alpha,\beta)}E_A(T)
$$
as $\alpha,\beta\To 0$.
\item Parts~(\ref{m.pdfr.adm}) and (\ref{m.pfdr.opt}) still hold if $J\To\infty$ as $\alpha, \beta\To 0$ such that $J=o(|\log(\alpha\wedge \beta)|^{1/4})$, \eqref{KL.inf.bound}-\eqref{KL.var.bound} hold, and $m\in[J-1]$ varies arbitrarily with $\alpha, \beta$.
\end{enumerate}
\end{corollary}

\subsection{Other multiple testing error metrics}\label{sec:gap.other}
Theorem~\ref{thm:gap} applies to any multiple testing error metrics which satisfy the theorem's conditions~(\ref{m.mte<}) and (\ref{m.mte>}). In addition to FDR/FNR and pFDR/pFNR these include the \textit{per-comparison error rate (PCER)} $E(V/J)$ as defined by \citet{Benjamini95}, the \textit{false positive rate} $E(V/m)$ \citep[e.g.,][]{Burke88}, \textit{per-family error rate\footnote{This metric is not constrained to take values in $[0,1]$ as we assumed above, but the theory there can be modified to allow metrics to take any nonnegative values with only notational changes. Alternatively, one could think of standardizing the false positive rate by dividing by $m$ or $J$, leading to the other two metrics mentioned.} (PFER)} $EV$ which appears in social and behavioral science research~\cite[e.g.,][]{Frane15,Keselman15}, and their type~2 analogs. These type~1 metrics are all proportional to $EV$ and for the gap rule,
\begin{equation}\label{gap.EV<}
EV=E(V\bm{1}\{V\ge 1\})\le m\cdot P(V\ge 1)=m\cdot\fwer_1,
\end{equation}
and for any procedure,
\begin{equation}\label{gap.EV>}
EV=E(V\bm{1}\{V\ge 1\}) \ge 1\cdot P(V\ge 1)= \fwer_1.
\end{equation}
Similar statements apply to the type~2 versions.

\section{Bounds on the number of signals}\label{sec:bds}
In this section we consider asymptotically optimal procedures for scenarios in which the number of signals is known to be between two given values $0\le\ell<u\le J$,  the $\ell=u$ case having already been considered in Section~\ref{sec:m}. That is, we consider asymptotic optimality in signal classes of the form $\mP_{\ell,u}:=\{A\subseteq [J]: \ell\le|A|\le u\}$ for such $\ell, u$. Mirroring our results for when the number of signals is known exactly, we show that the sequential procedures that are asymptotically optimal for FWE control are also asymptotically optimal (with modified critical values) for control of any multiple testing error metric whose  type~1 and 2 versions are bounded between constant multiples of the corresponding $\fwer_i$ in certain situations, made precise in conditions (\ref{lu.mte<})-(\ref{lu.mte>}) of Theorem~\ref{thm:GI}.

Given such bounds $\ell$ and $u$, \citet{Song17} defined the following \textit{gap-intersection rule}~$(T_{GI}, D_{GI})$, depending on four positive thresholds $a,b,c,d$, in terms of the stopping times
\begin{align}
\tau_1 &:= \inf\{n \geq 1:\; \lambda^{(\ell+1)}(n) \leq -a,\;  \lambda^{(\ell)}(n)\ -  \lambda^{(\ell + 1)}(n) \ge  c\},\label{tau1}\\
\tau_2 &:= \inf\{n \geq 1:\; \ell \le p(n) \leq u\qmq{and}   \lambda^{(j)}(n) \notin (-a, b) \qm{for all $j\in[J]$}\},\label{tau2}\\
\tau_3 &:= \inf\{n \geq 1:\; \lambda^{(u)}(n) \geq b,\;  \lambda^{(u)}(n)- \lambda^{(u + 1)}(n) \geq d\}.\label{tau3}
\end{align}
In \eqref{tau1}-\eqref{tau3} we set $\lambda^{(0)}(n)=-\infty$ and $\lambda^{(J)}(n)=\infty$ for all $n$ to handle the cases $\ell=0$ or $u=J$. Finally, the gap-intersection rule~$(T_{GI}, D_{GI})$ is defined as 
$$T_{GI} := \min\{\tau_1, \tau_2, \tau_3\}\qmq{and} D_{GI} := \{ i_1(T_{GI}),...,i_{p'}(T_{GI})\},$$
where $p' := (p(T_{GI})\vee \ell) \wedge u$ is the value in  $\{\ell, \ell+1,\ldots,u\}$ closest to $p(T_{GI})$. \citet{Song17} describe $\tau_2$ as similar to De and Baron's \citeyearpar{De12} ``intersection rule,'' which requires only the second condition in \eqref{tau2}, but modified to incorporate the prior information on the number of signals by requiring $\ell \le p(\tau_2) \leq u$. The stopping times $\tau_1$ and $\tau_3$ provide needed additional efficiency when the number of signals equals $\ell$ or $u$.

\subsection{Main result and its application to FDR/FNR and pFDR/pFNR control}
\begin{theorem}\label{thm:GI} 
Fix integers $0\le\ell<u\le J$ and let $(T_{GI},D_{GI})$ denote the gap-intersection rule with bounds $\ell, u$ on the number of signals and thresholds~$a,b,c,d>0$. Let MTE be multiple testing error metric such that:
\begin{enumerate}[(i)]
\item\label{lu.mte<}  there is a constant~$C_1$ such that
\begin{equation}\label{lu.mte<disp}
\mte_{i,A}(T_{GI},D_{GI}) \le C_1\cdot \fwer_{i,A}(T_{GI},D_{GI})
\end{equation}
for $i=1$ and $2$, for all $A\in\mP_{\ell,u}$, and for all thresholds~$a,b,c,d>0$, and
\item\label{lu.mte>} there is a constant~$C_2$ such that
\begin{equation}\label{lu.mte>ineq}
\mte_{i,A}(T,D) \ge C_2\cdot \fwer_{i,A}(T,D)
\end{equation}
for $i=1$ and $2$, for all $A\in\mP_{\ell,u}$, and for all procedures $(T,D)$.
\end{enumerate}
Given $\alpha,\beta\in(0,1)$ let $(T_{GI}^*,D_{GI}^*)$ denote the gap rule with bounds $\ell,u$ on the number of signals, and  thresholds 
\begin{align*}
&a = |\log(\beta/C_1)| +\ \log J, &&b = |\log(\alpha/C_1)| +\ \log J,\\
&c = |\log(\alpha/C_1)| +\ \log((J - \ell)J), &&d = |\log(\beta/C_1)| +\ \log(u J).
\end{align*}
Then the following hold.
\begin{enumerate}[(1)]
\item\label{lu.adm} Under the assumption that the streams $X^1,\ldots, X^J$ are independent, $(T_{GI}^*,D_{GI}^*)$ is admissible for MTE control. That is,  
\begin{equation}\label{lu.T*inD}
(T_{GI}^*,D_{GI}^*)\in \Delta_{\mP_{\ell,u}}^{\tiny{\mte}}(\alpha,\beta).
\end{equation}
\item\label{lu.opt} Under the additional assumption that each stream $X^j=(X_1^j, X_2^j,\ldots)$ is made up of i.i.d.\ random variables~$X_1^j, X_2^j,\ldots$, the procedure $(T_{GI}^*,D_{GI}^*)$ is asymptotically optimal for MTE control with respect to class~$\mP_{\ell,u}$. That is, for all $A \in \mP_{\ell,u}$,
\begin{align}
E_A(T_{GI}^*) & \sim \inf_{(T, D)\in\Delta_{\mP_{\ell,u}}^{\tiny{\mte}}(\alpha,\beta)} E_A(T)\nonumber\\ 
&\sim \begin{cases}
\max\{ |\log\beta|/\eta_0^A, |\log\alpha|/(\eta_0^A + \eta_1^A)\}, &\mbox{if $|A| = \ell$}\\
\max\{ |\log\beta|/\eta_0^A, |\log\alpha|/\eta_1^A\}, &\mbox{if $\ell < |A| < u$}\\
\max\{ |\log\alpha|/\eta_1^A, |\log\beta|/(\eta_0^A + \eta_1^A)\}, &\mbox{if $|A| = u$}
\end{cases}\label{lu.kappa}
\end{align}
as $\alpha,\beta\To 0$.
\item\label{lu.asymp} Parts~(\ref{lu.adm}) and (\ref{lu.opt}) still hold if $J, m, \ell, u, C_1$, and $C_2$ are allowed to vary with $\alpha, \beta$ as long as \eqref{m.J.inf}-\eqref{KL.var.bound} hold. In particular, the asymptotic optimality in part~(\ref{lu.opt}) still holds if $J, C_1\To\infty$ and $C_2\To 0$ as $\alpha,\beta\To 0$ as long as \eqref{m.J.inf}-\eqref{KL.var.bound} hold, with $m\in[J-1]$ and $0\le\ell<u\le J$ allowed to vary arbitrarily.
\end{enumerate}
\end{theorem}

\bigskip

\noindent\textbf{Proof.} For part~(\ref{lu.adm}), fix arbitrary $A\in\mP_{\ell,u}$. \citet[][Theorem~3.2]{Song17} establish that
\begin{equation*}
\fwer_{1,A}(T_{GI}^*,D_{GI}^*)\le \alpha/C_1\qmq{and}\fwer_{2,A}(T_{GI}^*,D_{GI}^*)\le \beta/C_1.
\end{equation*}
Applying \eqref{lu.mte<disp} yields
$$\mte_{1,A}(T_{GI}^*,D_{GI}^*)\le C_1\cdot \fwer_{1,A}(T_{GI}^*,D_{GI}^*)\le C_1(\alpha/C_1)=\alpha,$$
and a similar argument shows that $\mte_{2,A}(T_{GI}^*,D_{GI}^*)\le\beta$, establishing \eqref{lu.T*inD}.

For part~(\ref{lu.opt}), fix arbitrary $A\in\mP_{\ell,u}$ and consider $\alpha,\beta\To 0$. Letting $\kappa(\alpha,\beta)\To\infty$ denote the corresponding expression in \eqref{lu.kappa},  it is not hard to see that 
\begin{equation}\label{lu.k/C}
\kappa(\alpha/C_1,\beta/C_1)=\kappa(\alpha,\beta) +O(\log C_1),
\end{equation}
 which can be verified for each of the three cases.  
For the second case of \eqref{lu.kappa}, \citet[][Theorem~5.5]{Song17} establish that
\begin{equation}
E_A(T_{GI}^*)\le \kappa(\alpha/C_1,\beta/C_1)+O(J\sqrt{a\vee b})=  \kappa(\alpha,\beta)+O(\log C_1)+O(J\sqrt{\kappa(\alpha,\beta)})\sim \kappa(\alpha,\beta).\label{ETGI.low}
\end{equation}
And similar arguments show that in the first case of \eqref{lu.kappa},
\begin{equation}
E_A(T_{GI}^*)\le \kappa(\alpha/C_1,\beta/C_1)+O(J\sqrt{a\vee c})=  \kappa(\alpha,\beta)+O(\log C_1)+O(J\sqrt{\kappa(\alpha,\beta)})\sim \kappa(\alpha,\beta),\label{ETGI.low1}
\end{equation}
and in the third case of \eqref{lu.kappa},
\begin{equation}
E_A(T_{GI}^*)\le \kappa(\alpha/C_1,\beta/C_1)+O(J\sqrt{b\vee d})=  \kappa(\alpha,\beta)+O(\log C_1)+O(J\sqrt{\kappa(\alpha,\beta)})\sim \kappa(\alpha,\beta).\label{ETGI.low3}
\end{equation}

To establish that $\kappa(\alpha,\beta)$ is also an asymptotic lower bound in $\Delta_{\mP_{\ell,u}}^{\tiny{\mte}}(\alpha,\beta)$, it follows from \eqref{lu.mte>ineq} that $\Delta_{\mP_{\ell,u}}^{\tiny{\mte}}(\alpha,\beta)\subseteq \Delta_{\mP_{\ell,u}}^{\tiny{\fwer}}(\alpha/C_2, \beta/C_2)$, thus
$$\inf_{(T, D)\in \Delta_{\mP_{\ell,u}}^{\tiny{\mte}}(\alpha,\beta)}E_A(T) \ge \inf_{(T, D)\in \Delta_{\mP_{\ell,u}}^{\tiny{\fwer}}(\alpha/C_2, \beta/C_2)}E_A(T).$$
\citet[][Theorem~5.5]{Song17} establish that the latter is greater than or equal to
$$\kappa(\alpha/C_2,\beta/C_2)(1+o(1))= \kappa(\alpha,\beta)+O(|\log C_2|),$$ by arguments similar to \eqref{lu.k/C}. Thus
\begin{equation}\label{lu.low.equiv}
\inf_{(T, D)\in \Delta_{\mP_{\ell,u}}^{\tiny{\mte}}(\alpha,\beta)}E_A(T) \ge \kappa(\alpha,\beta)+O(|\log C_2|)\sim \kappa(\alpha,\beta).
\end{equation}
and combining this with \eqref{ETGI.low} gives the desired result.

The proof of part~(\ref{lu.asymp}) is similar to the corresponding part in Theorem~\ref{thm:gap} since the equivalences in \eqref{ETGI.low}-\eqref{lu.low.equiv} still hold when the parameters and constants are allowed to vary with $\alpha, \beta$ in the way described. \qed

\subsubsection{FDR/FNR control under bounds on the number of signals}

The application of Theorem~\ref{thm:GI} to  FDR/FNR is again immediate from the bounds \eqref{fdr<fwe}-\eqref{fnr<fwe}, from which we see that  condition~(\ref{lu.mte<}) of the theorem holds with $C_1=1$ and condition~(\ref{lu.mte>}) holds with $C_2=1/J$, yielding the following corollary. And letting $J\To \infty$ such that $J=o(|\log(\alpha\wedge \beta)|^{1/4})$ satisfies the requirements in \eqref{m.J.inf} of both $J$ and $C_2=1/J$.

\begin{corollary}\label{cor:GI.fdr} 
Given $\alpha,\beta\in(0,1)$ and integers $0\le\ell<u\le J$, let $(T_{GI}^*,D_{GI}^*)$ denote the gap-intersection rule with bounds $\ell,u$ on the number of signals, and  thresholds 
\begin{align*}
&a = |\log\beta| +\ \log J, &&b = |\log \alpha| +\ \log J,\\
&c = |\log \alpha | +\ \log((J - \ell)J), &&d = |\log\beta| +\ \log(u J).
\end{align*}
Then the following hold.
\begin{enumerate}[(1)]
\item\label{lu.fdr.adm} Under the assumption that the streams $X^1,\ldots, X^J$ are independent, $(T_{GI}^*,D_{GI}^*)$ is admissible for FDR/FNR control. That is,  
\begin{equation*}
(T_{GI}^*,D_{GI}^*)\in \Delta_{\mP_{\ell,u}}^{\tiny{\fdr}}(\alpha,\beta).
\end{equation*}
\item\label{lu.fdr.opt} Under the additional assumption that each stream $X^j=(X_1^j, X_2^j,\ldots)$ is made up of i.i.d.\ random variables~$X_1^j, X_2^j,\ldots$, the procedure $(T_{GI}^*,D_{GI}^*)$ is asymptotically optimal for FDR/FNR control with respect to class~$\mP_{\ell,u}$. That is, for all $A \in \mP_{\ell,u}$,
\begin{align*}
E_A(T_{GI}^*) & \sim \inf_{(T, D)\in\Delta_{\mP_{\ell,u}}^{\tiny{\fdr}}(\alpha,\beta)} E_A(T)\\ 
&\sim \begin{cases}
\max\{ |\log\beta|/\eta_0^A, |\log\alpha|/(\eta_0^A + \eta_1^A)\}, &\mbox{if $|A| = \ell$}\\
\max\{ |\log\beta|/\eta_0^A, |\log\alpha|/\eta_1^A\}, &\mbox{if $\ell < |A| < u$}\\
\max\{ |\log\alpha|/\eta_1^A, |\log\beta|/(\eta_0^A + \eta_1^A)\}, &\mbox{if $|A| = u$}
\end{cases}
\end{align*}
as $\alpha,\beta\To 0$.
\item Parts~(\ref{lu.fdr.adm}) and (\ref{lu.fdr.opt}) still hold if $J\To\infty$ as $\alpha, \beta\To 0$ such that $J=o(|\log(\alpha\wedge \beta)|^{1/4})$, \eqref{KL.inf.bound}-\eqref{KL.var.bound} hold, and $m\in[J-1]$ and $0\le\ell<u\le J$ vary arbitrarily with $\alpha, \beta$.
\end{enumerate}
\end{corollary}

\subsubsection{pFDR/pFNR control  under bounds on the number of signals}\label{sec:GI.pfdr}
In applying Theorem~\ref{thm:GI} to  pFDR/pFNR, some care must be taken when considering condition~(\ref{lu.mte<}) of the theorem since $\pfdr$ and $\pfnr$ cannot be universally bounded above by a constant multiple of $\fwer_1$ and $\fwer_2$, respectively, as in \eqref{pFDR<}. Although condition~(\ref{lu.mte<}) only requires such bounds to hold for the gap-intersection rule, these may still fail when the lower bound is $\ell=0$ and $P(R\ge 1)$ is close to zero, or when the upper bound is $u=J$ and $P(J-R\ge 1)$ is close to zero.  For this reason, in the following corollary which applies Theorem~\ref{thm:GI} to  pFDR/pFNR control, we restrict attention to prior bounds $\ell\ge 1$ and $u\le J-1$, under which the gap-intersection rule satisfies $P(R=0)=P(R=J)=0$ and thus condition~(\ref{lu.mte<}) holds with $C_1=1$ by \eqref{pFDR<}. We note that problems with pFDR and pFNR when $P(R\ge 1)$ and $P(J-R\ge 1)$, respectively, are close to zero are not unique to our setup here, and these quantities are of course undefined when these probabilities equal zero; see Section~\ref{sec:disc} for more discussion of this topic. The  needed lower bounds in condition~(\ref{lu.mte>}) of the theorem are again provided by the universal bounds \eqref{pFD/FW>}-\eqref{pFN/FW>} with $C_2=1/J$.

\begin{corollary}\label{cor:GI.pfdr} 
Given $\alpha,\beta\in(0,1)$ and integers $1\le\ell<u\le J-1$, let $(T_{GI}^*,D_{GI}^*)$ denote the gap-intersection rule with bounds $\ell,u$ on the number of signals, and  thresholds 
\begin{align*}
&a = |\log\beta| +\ \log J, &&b = |\log \alpha| +\ \log J,\\
&c = |\log \alpha | +\ \log((J - \ell)J), &&d = |\log\beta| +\ \log(u J).
\end{align*}
Then the following hold.
\begin{enumerate}[(1)]
\item\label{lu.pfdr.adm} Under the assumption that the streams $X^1,\ldots, X^J$ are independent, $(T_{GI}^*,D_{GI}^*)$ is admissible for pFDR/pFNR control. That is,  
\begin{equation*}
(T_{GI}^*,D_{GI}^*)\in \Delta_{\mP_{\ell,u}}^{\tiny{\pfdr}}(\alpha,\beta).
\end{equation*}
\item\label{lu.pfdr.opt} Under the additional assumption that each stream $X^j=(X_1^j, X_2^j,\ldots)$ is made up of i.i.d.\ random variables~$X_1^j, X_2^j,\ldots$, the procedure $(T_{GI}^*,D_{GI}^*)$ is asymptotically optimal for pFDR/pFNR control with respect to class~$\mP_{\ell,u}$. That is, for all $A \in \mP_{\ell,u}$,
\begin{align*}
E_A(T_{GI}^*) & \sim \inf_{(T, D)\in\Delta_{\mP_{\ell,u}}^{\tiny{\pfdr}}(\alpha,\beta)} E_A(T)\\ 
&\sim \begin{cases}
\max\{ |\log\beta|/\eta_0^A, |\log\alpha|/(\eta_0^A + \eta_1^A)\}, &\mbox{if $|A| = \ell$}\\
\max\{ |\log\beta|/\eta_0^A, |\log\alpha|/\eta_1^A\}, &\mbox{if $\ell < |A| < u$}\\
\max\{ |\log\alpha|/\eta_1^A, |\log\beta|/(\eta_0^A + \eta_1^A)\}, &\mbox{if $|A| = u$}
\end{cases}
\end{align*}
as $\alpha,\beta\To 0$.
\item Parts~(\ref{lu.pfdr.adm}) and (\ref{lu.pfdr.opt}) still hold if $J\To\infty$ as $\alpha, \beta\To 0$ such that $J=o(|\log(\alpha\wedge \beta)|^{1/4})$, \eqref{KL.inf.bound}-\eqref{KL.var.bound} hold, and $m\in[J-1]$ and $0<\ell<u< J$ vary arbitrarily with $\alpha, \beta$.
\end{enumerate}
\end{corollary}

\subsection{Other multiple testing error metrics}\label{sec:GI.other}
In addition to FDR/FNR and pFDR/pFNR, Theorem~\ref{thm:GI} applies to any multiple testing error metrics which satisfy the theorem's conditions including the metrics mentioned in Section~\ref{sec:gap.other}. The argument is similar, with the upper bound~\eqref{gap.EV<} being replaced by $u\cdot\fwer_1$ for the gap-intersection rule.

\section{Simulation study}\label{sec:sim}
In this section we study the non-asymptotic performance of the sequential gap rule for FDR/FNR control   as well as comparable fixed-sample procedures based on the Benjamini-Hochberg \citeyearpar{Benjamini95} procedure (BH) through a simulation study in the setting of Section~\ref{sec:m} where the number~$m$ of signals is known exactly. We consider $J$ independent streams $X_1^j, X_2^j,\ldots$, $j\in[J]$, of i.i.d.\ $N(\mu,1)$ data with hypotheses $H_0^j: \mu=0$ vs.\ $H_1^j: \mu=1/2$, $j\in[J]$. Table~\ref{tab:10} contains the operating characteristics of three procedures under this setup with $J=10$ and all possible values of $m\in [J-1]$ in the table's rows.  The columns under the heading Gap Rule in Table~\ref{tab:10} describe the performance of the gap rule as defined in \eqref{gapT}-\eqref{gapD} with values of the threshold~$c$ listed in the table. In order to study the non-asymptotic performance of this procedure, the asymptotic value of $c$  in Corollary~\ref{cor:gap.fdr} was dispensed with and for each $m$, a value of $c$ was determined by Monte Carlo simulation such that both the achieved FDR and FNR are close to, but no larger than, the nominal values  $\alpha=\beta=.05$. We note that this same approach is available to users in practice since $m$ and the simple hypotheses $H_0^j, H_1^j$ are all known and thus the achieved FDR and FNR can be estimated to arbitrary accuracy via Monte Carlo before gathering any data. For the gap rule with $c$ determined in this way, the expected stopping time $ET$ and achieved FDR and FNR, estimated from 10,000 Monte Carlo replications, are given in the table with their standard errors given in parentheses.

\begin{sidewaystable}[htp]
\caption{Operating characteristics of sequential gap rule and fixed-sample BH and BH$_m$ procedures in simulation study as described in Section~\ref{sec:sim} with $J=10$ data streams.}
\begin{tabular}{|c|c|c|c|c|c|c|c|c|c|c|}
\hline
  & \multicolumn{4}{c|}{Gap Rule}              & \multicolumn{3}{c|}{BH}             & \multicolumn{3}{c|}{BH$_m$}          \\ \hline
$m$ & $c$   & $ET$   & FDR (\%)        & FNR (\%)       &$n$ (Savings) & FDR  (\%)      & FNR   (\%)     &$n$ (Savings)& FDR  (\%)      & FNR    (\%)    \\ \hline
1 & 3.5 & 29.0 (0.15) & 4.30 (0.20) & 0.48 (0.02) & 70 (59\%)  & 4.49 (0.15) & 0.61 (0.02) & 50 (42\%)  & 4.54 (0.21) & 0.50 (0.02) \\ \hline
2 & 2.9 & 31.6 (0.14) & 4.60 (0.15) & 1.15 (0.04) & 60 (47\%)  & 3.97 (0.12) & 1.50 (0.04) & 46 (31\%)  & 4.74 (0.15) & 1.18 (0.04) \\ \hline
3 & 2.6 & 31.7 (0.14) & 4.75 (0.12) & 2.04 (0.05) & 59 (46\%)  & 3.55 (0.09) & 2.05 (0.05) & 45 (30\%)  & 4.40 (0.11) & 1.88 (0.05) \\ \hline
4 & 2.3 & 30.2 (0.13) & 4.59 (0.10) & 3.06 (0.07) & 54 (44\%)  & 2.84 (0.08) & 3.24 (0.07) & 40 (25\%)  & 4.80 (0.10) & 3.20 (0.06) \\ \hline
5 & 2.1 & 28.7 (0.12) & 4.66 (0.09) & 4.66 (0.09) & 52 (45\%)  & 2.55 (0.07) & 4.67 (0.08) & 37 (22\%)  & 4.75 (0.09) & 4.75 (0.09) \\ \hline
6 & 2.3 & 30.5 (0.13) & 3.18 (0.07) & 4.77 (0.10) & 54 (44\%)  & 2.06 (0.05) & 4.53 (0.09) & 40 (24\%)  & 3.32 (0.07) & 4.99 (0.10) \\ \hline
7 & 2.5 & 30.8 (0.13) & 2.14 (0.05) & 4.90 (0.12) & 56 (45\%)  & 1.50 (0.04) & 4.91 (0.11) & 43 (28\%)  & 2.10 (0.05) & 4.90 (0.12) \\ \hline
8 & 2.8 & 30.7 (0.14) & 1.22 (0.04) & 4.89 (0.15) & 60 (49\%)  & 0.96 (0.03) & 4.92 (0.13) & 45 (32\%)  & 1.31 (0.04) & 5.24 (0.15) \\ \hline
9 & 3.4 & 28.5 (0.15) & 0.49 (0.02) & 4.39 (0.20) & 65 (56\%)  & 0.50 (0.02) & 4.77 (0.15) & 50 (43\%)  & 0.48 (0.02) & 4.33 (0.20) \\ \hline
\end{tabular}
\label{tab:10}
\end{sidewaystable}

A natural competitor for the gap rule is the fixed-sample BH procedure. In implementing the BH procedure  two parameters must be chosen: the procedure's nominal FDR control level~$\alpha$ and its fixed sample size, denoted by $n$ in Table~\ref{tab:10}.  In order to have an even-handed comparison, the same nominal level $\alpha=.05$ as the gap rule was used, and for each $m$ the BH procedure's sample size~$n$ was then  chosen so that its achieved FNR was as close as possible to that of the gap rule. The columns under the heading BH in Table~\ref{tab:10} contain the operating characteristics of the BH procedure with parameters chosen in this way, where again the FDR and FNR are estimated from 10,000 Monte Carlo replications with standard errors given in parentheses. The first of these columns contains  the fixed sample size~$n$ as well as the expected savings~$1-ET/n$ in sample size  of the sequential gap rule over the fixed-sample rule. 

Although the BH procedure as implemented in the previous paragraph does make some use of knowledge of the number~$m$ of signals through the choice of its fixed sample size~$n$ for each $m$, the decision rule itself does not make explicit use of $m$ and may not necessarily reject exactly $m$ null hypotheses. In order compare with a fixed-sample procedure that does make explicit use of $m$, like the gap rule, we also implemented a fixed-sample procedure  (denoted BH$_m$) that always rejects exactly $m$ null hypotheses by sampling $n$ observations from each stream and rejecting the $m$ null hypotheses with the smallest $p$-values. Such a procedure makes no use of a nominal level~$\alpha$, so we implemented BH$_m$ by choosing $n$ so that its achieved FDR and FNR are close to, but no larger than, the nominal levels $\alpha=\beta=.05$, similar to how the gap rule's threshold $c$ was determined.  The last three columns of Table~\ref{tab:10}, under the heading  BH$_m$, give the operating characteristics of this procedure, including the savings in sample size achieved by the gap rule.

From the Savings columns in Table~\ref{tab:10} we see that the sequential gap rule provides dramatic efficiency gains, in terms of average sample size, relative to the BH and BH$_m$ procedures in this setting. The savings is in the 40\%-60\% range versus BH, and less so but still substantial in the 20\%-45\% range versus BH$_m$, which makes more explicit use of the true number of signals~$m$ than the BH procedure.  The savings in sample size is largest for values of $m$ near $0$ and $J$, and decreases for $m$ near $J/2$.  The achieved FDR and FNR values are comparable among the three procedures.

\begin{sidewaystable}[htp]
\caption{Operating characteristics of sequential gap rule and fixed-sample BH and BH$_m$ procedures in simulation study as described in Section~\ref{sec:sim} with $J=100$ data streams.}
\begin{tabular}{|c|c|c|c|c|c|c|c|c|c|c|}
\hline
  & \multicolumn{4}{c|}{Gap Rule}              & \multicolumn{3}{c|}{BH}             & \multicolumn{3}{c|}{BH$_m$}          \\ \hline
$m$ & $c$   & $ET$   & FDR (\%)       & FNR   (\%)     &$n$ (Savings) & FDR   (\%)     & FNR    (\%)    &$n$ (Savings)& FDR   (\%)     & FNR  (\%)      \\ \hline
1  & 3.9 & 48.8 (0.21) & 4.43 (0.21) & 0.04 (0.00) & 90 (46\%)  & 4.77 (0.16) & 0.08 (0.00) & 77 (37\%)  & 4.74 (0.21) & 0.05 (0.00) \\ \hline
10 & 1.9 & 61.0 (0.16) & 4.65 (0.06) & 0.52 (0.01) & 70 (13\%)  & 4.52 (0.06) & 0.63 (0.01) & 68 (10\%)  & 4.73 (0.06) & 0.53 (0.01) \\ \hline
20 & 1.3 & 57.3 (0.14) & 4.76 (0.04) & 1.19 (0.01) & 65 (12\%)  & 3.94 (0.04) & 1.17 (0.01) & 62 (8\%)   & 4.57 (0.04) & 1.14 (0.01) \\ \hline
30 & 1.0   & 52.8 (0.12) & 4.70 (0.03) & 2.01 (0.01) & 60 (12\%)  & 3.50 (0.03) & 2.01 (0.02) & 57 (7\%)   & 4.81 (0.02) & 3.21 (0.02) \\ \hline
40 & 0.8 & 48.0 (0.11) & 4.74 (0.03) & 3.16 (0.02) & 56 (14\%)  & 3.00 (0.03) & 3.22 (0.02) & 50 (3\%)   & 4.81 (0.02) & 3.20 (0.02) \\ \hline
50 & 0.7 & 45.1 (0.10) & 4.47 (0.03) & 4.47 (0.03) & 53 (15\%)  & 2.53 (0.02) & 4.90 (0.03) & 47 (4\%)   & 4.39 (0.02) & 4.39 (0.02) \\ \hline
60 & 0.8 & 48.2 (0.11) & 3.19 (0.02) & 4.79 (0.03) & 56 (14\%)  & 2.02 (0.02) & 4.94 (0.03) & 50 (3\%)   & 3.17 (0.02) & 4.76 (0.02) \\ \hline
70 & 1.0   & 52.8 (0.12) & 2.03 (0.01) & 4.74 (0.03) & 60 (12\%)  & 1.52 (0.01) & 4.74 (0.04) & 57 (7\%)   & 2.00 (0.01) & 4.59 (0.03) \\ \hline
80 & 1.3 & 57.0 (0.13) & 1.20 (0.01) & 4.78 (0.04) & 64 (11\%)  & 1.00 (0.01) & 5.00 (0.05) & 63 (10\%)  & 1.09 (0.02) & 4.38 (0.04) \\ \hline
90 & 1.9 & 61.8 (0.16) & 0.51 (0.01) & 4.63 (0.06) & 72 (14\%)  & 0.50 (0.01) & 4.87 (0.06) & 71 (13\%)  & 0.47 (0.01) & 4.24 (0.06) \\ \hline
99 & 3.9 & 48.7 (0.21) & 0.04 (0.00) & 4.10 (0.20) & 90 (46\%)  & 0.05 (0.00) & 5.62 (0.17) & 79 (38\%)  & 0.04 (0.00) & 4.13 (0.20) \\ \hline
\end{tabular}
\label{tab:100}
\end{sidewaystable}

Table~\ref{tab:100} contains the operating characteristics of these procedures with parameters chosen in the analogous way for the $J=100$ data stream version of the same setup. The three procedures have a similar relationship to each other as in the $J=10$ setting, however the savings in sample size is less pronounced for this larger value of $J$.

\section{Discussion}\label{sec:disc}
\subsection*{Summary and extensions}
We have shown that sequential procedures proposed and shown to be asymptotically optimal by \citet{Song17} for FWE control can be made by modification of their critical values to be asymptotically optimal for control of other metrics, including FDR/FNR and pFDR/pFNR, in the setting of a known number of signals, or bounds on the true number of signals. One interpretation of these results is that first order optimality is not fine grain enough to distinguish between $\fwer$, $\fdr$, $\pfdr$, and other metrics satisfying the main theorems, \ref{thm:gap} and \ref{thm:GI}, however it remains an open question whether optimal procedures for control of these metrics \emph{must} look different asymptotically.

The boundedness conditions (i)-(ii)  in Theorems~\ref{thm:gap} and \ref{thm:GI} are required to hold for all values of the nominal error probabilities~$\alpha$ and $\beta$, but this is just to obtain admissibility of the gap and gap-intersection rules (Part~1 of Theorems~\ref{thm:gap} and \ref{thm:GI}, respectively) for all $\alpha$ and $\beta$, and is not required for asymptotic optimality  (Part~2 of the theorems) which only need to consider $\alpha, \beta$ near $0$.  Thus, the optimality results can be further extended to multiple testing error metrics that only satisfy the boundedness conditions (i)-(ii) for small $\alpha, \beta$.

\subsection*{Exclusion of the $\ell=0$, $u=J$ cases for pFDR/pFNR control}\label{sec:no.0J}
In Section~\ref{sec:GI.pfdr} we have ruled out the cases of the number of signals equal to $0$ or $J$ for pFDR/pFNR control based on the reasoning that either of these metrics are not defined for procedures or scenarios in which $P(R=0)=1$ or $P(R=J)=1$, which are plausible when no, or all, streams contain signals. Focusing on the case of no signals (with similar remarks applying to the other case), the event $R=0$ of no rejections has previously been recognized as a difficulty with utilizing pFDR in fixed-sample analyses, and so is not unique to the sequential sampling considered here.  Even in Storey's \citeyearpar{Storey02} original proposal for estimating pFDR, $R$ must be replaced by $R\vee 1$ in the denominator of his estimator to avoid a singularity on the event $R=0$, and the probability of this event must be bounded below, relying on independence and exact uniform distribution of the associated $p$-values. \citet[][Section~4.3]{Black04} points out that this approach introduces a bias into the estimate of pFDR and argues that an improved estimator should indeed be left undefined on $R=0$.  These arguments further support our choice to leave out the cases $\ell=0$ and $u=J$ in Corollary~\ref{cor:GI.pfdr}.

The other metrics mentioned in Section~\ref{sec:GI.other}, to which the general theorem applies, do not have this problem because of elementary bounds like \eqref{gap.EV<} (replacing $m$ by $u$) and \eqref{gap.EV>}. 

 A referee suggested an alternative approach to pFDR/pFNR control that would include the number of signals being $0$ or $J$ by introducing a hybrid error metric that equals FDR/FNR on these cases and pFDR/pFNR otherwise. Since the bounds $C_1=1$ and $C_2=1/J$ are the same for both FDR/FNR control in Corollary~\ref{cor:GI.fdr} and pFDR/pFNR control in Corollary~\ref{cor:GI.pfdr}, it immediately follows that the common gap intersection rule would be asymptotically optimal for controlling this hybrid metric.

\subsection*{Relationship to classification problems}

The problem addressed here may naturally be considered a classification problem, in which $J$ objects are each classified into one of two respective classes, i.e., the corresponding null and alternative hypotheses.  This is essentially the same setting  for which classification via FDR thresholding has been investigated for fixed-sample data by many authors including \citet{Abramovich06}, \citet{Bogdan08}, \citet{Donoho06}, \citet{Genovese02}, and \citet{Neuvial12}. FDR and related metrics have previously been considered as classifiers based on sequential data \citep[e.g.,][]{Bartroff18,Bartroff20} but the current paper appears to be the first time asymptotic optimality has been considered and achieved for these metrics.

The setup considered here accommodates more specialized classification problems as well.  For example, \textit{slippage problems} \citep[see][]{Dragalin99,Ferguson67,Mosteller48} consider $J$ independent populations, of which at most one  is in a non-null state, represented by the alternative hypothesis. In the fixed-sample setting, \citet{Ferguson67} considered a slippage problem in which both the null and alternative are known. \citet{Tartakovsky97} found minimax solutions for more general hypotheses, as did \citet{Dragalin99} in a sequential Bayesian setting. The current paper can address slippage problems by taking the bounds in Section~\ref{sec:bds} on the number of signals to be $\ell=0$ and $u=1$, since at most one stream has a signal, and allows the asymptotically optimal application of FDR/FNR or any other metric satisfying Theorem~\ref{thm:GI}. However, we note that the utility of FDR beyond $\fwer_1$ is inherently limited in slippage problems because the two metrics will coincide for any procedure which chooses either $\ell=0$ or $u=1$ signals. This is because, using the notation in the sections above, both the number~$V$ of falsely rejected nulls and the number~$R$ of rejected nulls can only take the values $0$ or $1$, hence
\begin{equation}\label{slip.fdr=fwe}
\fdr=E\left(\frac{V}{R\vee 1}\right)=E\left(\frac{V}{1}\right)=P(V=1)=P(V\ge 1)=\fwer_1.
\end{equation}
However, the same does not hold for $\fnr$ and $\fwer_2$ in this case hence, in spite of \eqref{slip.fdr=fwe}, the asymptotically optimal FDR/FNR control in the current paper is distinct from the $\fwer$ control of \citet{Song17} in slippage problems and its application here is novel, as well as  any other metrics satisfying Theorem~\ref{thm:GI}.

\section*{Acknowledgements} The authors are grateful for the  comments of two reviewers which improved the paper.  The majority of He's work was completed while she was a PhD student in the Department of Mathematics at the University of Southern California.


\begin{thebibliography}{}

\bibitem[Abramovich et~al., 2006]{Abramovich06}
Abramovich, F., Benjamini, Y., Donoho, D.~L., and Johnstone, I.~M. (2006).
\newblock Adapting to unknown sparsity by controlling the false discovery rate.
\newblock {\em Annals of Statistics}, 34(2):584--653.

\bibitem[Bartroff, 2018]{Bartroff18}
Bartroff, J. (2018).
\newblock Multiple hypothesis tests controlling generalized error rates for
  sequential data.
\newblock {\em Statistica Sinica}, 28:363--398.

\bibitem[Bartroff and Lai, 2010]{Bartroff10e}
Bartroff, J. and Lai, T.~L. (2010).
\newblock Multistage tests of multiple hypotheses.
\newblock {\em Communications in Statistics -- Theory and Methods (Special
  Issue Honoring M. Akahira, M. Aoshima, ed.)}, 39:1597--1607.

\bibitem[Bartroff and Song, 2014]{Bartroff14b}
Bartroff, J. and Song, J. (2014).
\newblock Sequential tests of multiple hypotheses controlling type {I} and {II}
  familywise error rates.
\newblock {\em Journal of Statistical Planning and Inference}, 153:100--114.

\bibitem[Bartroff and Song, 2015]{Bartroff15c}
Bartroff, J. and Song, J. (2015).
\newblock A rejection principle for sequential tests of multiple hypotheses
  controlling familywise error rates.
\newblock {\em Scandinavian Journal of Statistics}, 43:3--19.

\bibitem[Bartroff and Song, 2020]{Bartroff20}
Bartroff, J. and Song, J. (2020).
\newblock Sequential tests of multiple hypotheses controlling false discovery
  and nondiscovery rates.
\newblock {\em Sequential Analysis}.
\newblock To appear, preprint at \url{http://arxiv.org/abs/1311.3350}.

\bibitem[Benjamini and Hochberg, 1995]{Benjamini95}
Benjamini, Y. and Hochberg, Y. (1995).
\newblock Controlling the false discovery rate: {A} practical and powerful
  approach to multiple testing.
\newblock {\em Journal of the Royal Statistical Society, Series B:
  Methodological}, 57:289--300.

\bibitem[Black, 2004]{Black04}
Black, M. (2004).
\newblock A note on the adaptive control of false discovery rates.
\newblock {\em Journal of the Royal Statistical Society: Series B (Statistical
  Methodology)}, 66(2):297--304.

\bibitem[Bogdan et~al., 2008]{Bogdan08}
Bogdan, M., Ghosh, J.~K., and Tokdar, S.~T. (2008).
\newblock A comparison of the {B}enjamini-{H}ochberg procedure with some
  {B}ayesian rules for multiple testing.
\newblock In Balakrishnan, N., Pe\~na, E.~A., and Silvapulle, M.~J., editors,
  {\em Beyond Parametrics in Interdisciplinary Research: Festschrift in Honor
  of Professor Pranab K. Sen}, volume~1, pages 211--230, Beachwood, Ohio, USA.
  Institute of Mathematical Statistics.

\bibitem[Burke et~al., 1988]{Burke88}
Burke, D.~S., Brundage, J.~F., Redfield, R.~R., Damato, J.~J., Schable, C.~A.,
  Putman, P., Visintine, R., and Kim, H.~I. (1988).
\newblock Measurement of the false positive rate in a screening program for
  human immunodeficiency virus infections.
\newblock {\em New England Journal of Medicine}, 319(15):961--964.

\bibitem[Chen and Arias-Castro, 2017]{Chen17}
Chen, S. and Arias-Castro, E. (2017).
\newblock Sequential multiple testing.
\newblock ArXiV preprint, \url{http://arxiv.org/abs/1705.10190}.

\bibitem[De and Baron, 2012a]{De12}
De, S. and Baron, M. (2012a).
\newblock Sequential {B}onferroni methods for multiple hypothesis testing with
  strong control of family-wise error rates {I} and {II}.
\newblock {\em Sequential Analysis}, 31(2):238--262.

\bibitem[De and Baron, 2012b]{De12b}
De, S. and Baron, M. (2012b).
\newblock Step-up and step-down methods for testing multiple hypotheses in
  sequential experiments.
\newblock {\em Journal of Statistical Planning and Inference}, 142:2059--2070.

\bibitem[Donoho and Jin, 2006]{Donoho06}
Donoho, D. and Jin, J. (2006).
\newblock Asymptotic minimaxity of false discovery rate thresholding for sparse
  exponential data.
\newblock {\em Annals of Statistics}, 34(6):2980--3018.

\bibitem[Dragalin et~al., 1999]{Dragalin99}
Dragalin, V.~P., Tartakovsky, A.~G., and Veeravalli, V.~V. (1999).
\newblock Multihypothesis sequential probability ratio tests {I}: Asymptotic
  optimality.
\newblock {\em IEEE Transactions on Information Theory}, 45(7):2448--2461.

\bibitem[Ferguson, 1967]{Ferguson67}
Ferguson, T.~S. (1967).
\newblock {\em Mathematical Statistics: A Decision Theoretic Approach}.
\newblock Academic Press, New York.

\bibitem[Frane, 2015]{Frane15}
Frane, A.~V. (2015).
\newblock Are per-family type {I} error rates relevant in social and behavioral
  science?
\newblock {\em Journal of Modern Applied Statistical Methods}, 14(1).

\bibitem[Genovese and Wasserman, 2002]{Genovese02}
Genovese, C. and Wasserman, L. (2002).
\newblock Operating characteristics and extensions of the false discovery rate
  procedure.
\newblock {\em Journal of the Royal Statistical Society: Series B (Statistical
  Methodology)}, 64(3):499--517.

\bibitem[Gut, 2009]{Gut09}
Gut, A. (2009).
\newblock {\em Stopped Random Walks}.
\newblock Springer, New York.

\bibitem[Javanmard and Montanari, 2018]{Javanmard18}
Javanmard, A. and Montanari, A. (2018).
\newblock Online rules for control of false discovery rate and false discovery
  exceedance.
\newblock {\em The Annals of Statistics}, 46(2):526--554.

\bibitem[Jiang and Salzman, 2012]{Jiang12}
Jiang, H. and Salzman, J. (2012).
\newblock Statistical properties of an early stopping rule for resampling-based
  multiple testing.
\newblock {\em Biometrika}, 99(4):973--980.

\bibitem[Keselman, 2015]{Keselman15}
Keselman, H.~J. (2015).
\newblock Per family or familywise type {I} error control: ``{E}ether, eyether,
  neether, nyther, let's call the whole thing off!''.
\newblock {\em Journal of Modern Applied Statistical Methods}, 14(1):24--37.

\bibitem[Malloy and Nowak, 2014]{Malloy14}
Malloy, M.~L. and Nowak, R.~D. (2014).
\newblock Sequential testing for sparse recovery.
\newblock {\em {IEEE} Transactions on Information Theory}, 60(12):7862--7873.

\bibitem[Mei, 2008]{Mei08}
Mei, Y. (2008).
\newblock Asymptotic optimality theory for decentralized sequential hypothesis
  testing in sensor networks.
\newblock {\em {IEEE} Transactions on Information Theory}, 54(5):2072--2089.

\bibitem[Mosteller, 1948]{Mosteller48}
Mosteller, F. (1948).
\newblock A $k$-sample slippage test for an extreme population.
\newblock {\em Ann. Math. Statist.}, 19(1):58--65.

\bibitem[Neuvial and Roquain, 2012]{Neuvial12}
Neuvial, P. and Roquain, E. (2012).
\newblock On false discovery rate thresholding for classification under
  sparsity.
\newblock {\em Annals of Statistics}, 40(5):2572--2600.

\bibitem[Song and Fellouris, 2017]{Song17}
Song, Y. and Fellouris, G. (2017).
\newblock Asymptotically optimal, sequential, multiple testing procedures with
  prior information on the number of signals.
\newblock {\em Electron. J. Statist.}, 11(1):338--363.

\bibitem[Song and Fellouris, 2019]{Song19}
Song, Y. and Fellouris, G. (2019).
\newblock Sequential multiple testing with generalized error control: An
  asymptotic optimality theory.
\newblock {\em Ann. Statist.}, 47(3):1776--1803.

\bibitem[Storey, 2002]{Storey02}
Storey, J.~D. (2002).
\newblock A direct approach to false discovery rates.
\newblock {\em Journal of the Royal Statistical Society: Series B (Statistical
  Methodology)}, 64(3):479--498.

\bibitem[Tartakovsky, 1997]{Tartakovsky97}
Tartakovsky, A.~G. (1997).
\newblock Minimax invariant regret solution to the {$N$}-sample slippage
  problem.
\newblock {\em Mathematical Methods of Statistics}, 6(4):491--508.

\bibitem[Wegman and Marchette, 2003]{Wegman03}
Wegman, E.~J. and Marchette, D.~J. (2003).
\newblock On some techniques for streaming data: a case study of internet
  packet headers.
\newblock {\em Journal of Computational and Graphical Statistics},
  12(4):893--914.

\bibitem[Zrnic et~al., 2018]{Zrnic18}
Zrnic, T., Ramdas, A., and Jordan, M.~I. (2018).
\newblock Asynchronous online testing of multiple hypotheses.
\newblock ArXiV preprint, \url{http://arxiv.org/abs/1812.05068}.

\end{thebibliography}

\def\cprime{$'$}

\end{document}